\documentclass{appolb}
\usepackage{epsfig}

\newcommand{\lsim}{\raisebox{-0.13cm}{~\shortstack{$<$ \\[-0.07cm] $\sim$}}~}
\newcommand{\gsim}{\raisebox{-0.13cm}{~\shortstack{$>$ \\[-0.07cm] $\sim$}}~}
\newcommand{\beq}{\begin{eqnarray}}
\newcommand{\eeq}{\end{eqnarray}}

\begin{document}
\title{Higgs Boson Search at $e^+e^-$ and Photon Linear Colliders%
\thanks{Presented at the Photon Collider Workshop 2005, 5.-8.9.05, Kazimierz,
Poland}%
}
\author{M.M.~M\"uhlleitner
\address{Laboratoire d'Annecy-Le-Vieux de Physique Th\'eorique, LAPTH,
Annecy-Le-Vieux, France}
}
\maketitle
\begin{abstract}
The various search modes for the Higgs bosons of the Standard Model (SM) and
its Minimal Supersymmetric Extension (MSSM) at the International Linear 
Collider (ILC) will be summarized briefly. In particular, as a unique 
discovery mode the production of heavy neutral MSSM Higgs bosons for medium 
values of $\tan\beta$ in photon collisions will be presented. Furthermore, 
$\tau^+\tau^-$ fusion into MSSM Higgs bosons in the photon mode will be shown 
to give access to the mixing parameter $\tan\beta$ with a precision of better 
than 10\% for large values of this parameter.
\end{abstract}

\vspace*{-11.5cm}
\mbox{ }\hfill LAPTH-CONF-1134/05\\
\mbox{ }\hfill hep--ph/0512232\\

\vspace*{10cm}

\section{Introduction}
One of the major endeavours of high energy physics at future colliders is
the experimental test of the Higgs mechanism which allows to introduce 
standard particle masses without violating gauge symmetries. Four steps have 
to be taken \cite{Muhlleitner:2004mk}: First of all the Higgs boson(s) must 
be discovered. Next, the spin zero nature of the Higgs field can be verified 
through the determination of the Higgs boson quantum numbers. In the third 
step, by measuring its couplings to gauge bosons and fermions the 
proportionality to the masses of the respective particles as predicted by the 
Higgs mechanism can be checked. Finally, the triple and quartic Higgs 
self-couplings have to be determined in order to reconstruct the Higgs 
potential itself, responsible for the non-zero vacuum expectation value due to 
its specific form. In the following, the first step in this program will be 
summarized briefly. The discovery modes at the ILC and the Photon Linear 
Collider (PLC) will be discussed, complemented by a brief note on the 
extraction of $\tan\beta$ in $\tau^+\tau^-$ fusion at the PLC.

\section{Higgs boson search at the ILC}
\subsection{The SM Higgs boson}
\begin{figure}[h]
\begin{center}
\epsfig{figure=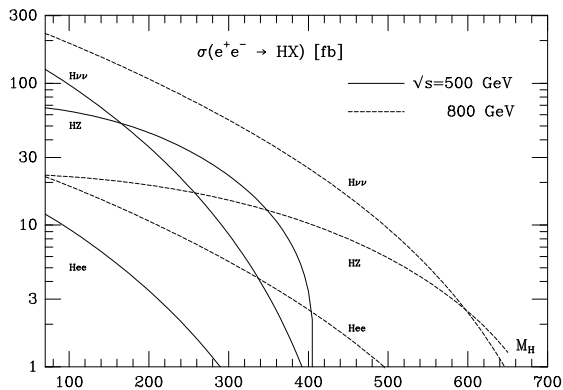,bbllx=125,bblly=449,bburx=477,bbury=692,height=5.1cm,clip=}
\caption{SM Higgs boson production processes as a function of the Higgs boson
mass for two typical collider energies, $\sqrt{s}=500,800$~GeV 
\cite{Aguilar-Saavedra:2001rg}.}
\label{smprod}
\end{center}
\end{figure}

The main SM Higgs boson production mechanisms are Higgs-strahlung, 
$e^+e^-\to ZH$ \cite{hrad} at lower energies and $WW$ fusion,
$e^+e^-\to H \bar{\nu}\nu$ \cite{fusion} at higher energies, 
{\it cf.}~Fig.\ref{smprod}. The full electroweak (EW) corrections at one loop 
have been calculated for both the Higgs-strahlung \cite{nlohrad,nloboth} and 
the fusion process \cite{nloboth,nlowfus}. They are of ${\cal O}(10\%)$. Since 
the recoiling $Z$ boson in Higgs-strahlung is monoenergetic at leading order 
the Higgs mass can be reconstructed independent of the Higgs boson decay. 
Combining recoil mass techniques and reconstruction of the Higgs decay 
products, the expected accuracy on $M_H$ is 40-80 MeV for intermediate mass 
Higgs bosons \cite{garcia}. 

\subsection{The MSSM Higgs bosons}
\begin{figure}[h]
\begin{center}
\epsfig{figure=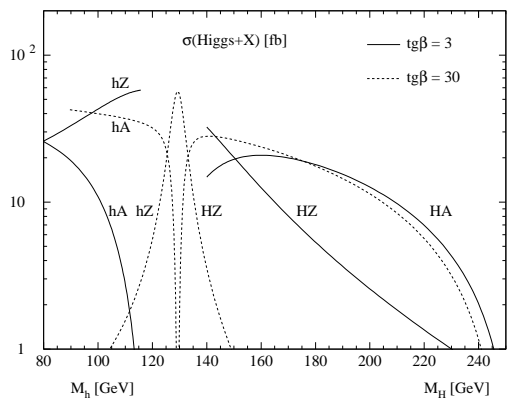,bbllx=57,bblly=125,bburx=537,bbury=738,width=5.2cm,clip=,angle=-90}
\caption{MSSM Higgs boson production as a function of $M_{h,H}$ for $\tan\beta=3,30$ \cite{Aguilar-Saavedra:2001rg}.}
\label{mssmprod}
\end{center}
\end{figure}

Supersymmetry and the requirement of an anomaly free theory require the 
introduction of two complex Higgs doublets in the MSSM, leading after 
EW symmetry breaking to 5 physical Higgs particles, 2 neutral CP-even $h,H$, 
one CP-odd $A$ and two charged bosons $H^\pm$. The neutral Higgs boson 
production mechanisms \cite{mssmprod} are Higgs-strahlung, $e^+e^- \to Z+h/H$, 
gauge boson fusion, $e^+e^-\to \bar{\nu}\nu/e^+e^- +h/H$, and associated 
production, $e^+e^-\to A+h/H$. Charged Higgs bosons are produced in pairs 
$e^+e^-\to H^+H^-$ or, if kinematically allowed, in top decays, $t\to H^+ b$. 
The production processes for $h,H$ as well as the Higgs-strahlung and 
associated production process are mutually complementary to each other 
{\it cf.}~Fig.~\ref{mssmprod}, coming either with $\sin^2(\beta-\alpha)$ or 
$\cos^2(\beta-\alpha)$ so that the lightest Higgs boson can always be 
discovered, its production cross section being large enough. All Higgs 
particles can be discovered at $\sqrt{s}=500$~GeV for masses below about 230 
GeV. If the Higgs decay modes are complicated or invisible, missing mass 
techniques allow their detection. Experimental studies have shown, that the 
$H,A$ masses can be measured with several hundred MeV accuracy in Higgs pair 
production far above the kinematical threshold \cite{raspereza}. The expected 
accuracy of $M_{H^\pm}$ is of order 1\% for $M_{H^\pm}=300$~GeV \cite{ferrari}.
\vspace*{-0.2cm}
%

An interesting MSSM parameter scenario is the intense-coupling regime, 
introduced in Ref.~\cite{icr}. All Higgs bosons are rather light and similar 
in mass, $M_h\sim M_H\sim M_A \lsim 130$~GeV, $M_{H^\pm}\lsim 150$~GeV. For 
large $\tan\beta$ values, one of the CP-even neutral Higgs bosons behaves as 
$A$ with large couplings to down-type fermions. The other behaves SM-like 
and couples strongly to $W,Z$ and top. Since the bosons are light, they are
in principle all accessible. The masses being rather close, several search 
channels have to be considered at the same time, further complicated by
sizable widths compared to the mass differences. In addition, the couplings 
can be significantly different from the SM or the MSSM decoupling limit. 
Experimental studies \cite{icrstudy} have shown that in a multichannel 
analysis the neutral Higgs masses can be extracted with an accuracy of 100-300 
MeV.

\section{Higgs bosons at the PLC}
\subsection{Heavy MSSM Higgs boson production in $\gamma\gamma$ collisions}
Heavy $H,A$ bosons may escape discovery at the Large Hadron Collider (LHC) for
intermediate values of $\tan\beta$ and are not accessible in the $e^+e^-$ 
mode of the ILC for masses above $\sqrt{s}/2$ \cite{Muhlleitner:2005pr}. 
The heavy $H,A$ appear as resonances in $\gamma\gamma$ collisions 
\cite{Muhlleitner:2001kw}. Therefore $\gamma\gamma \to H,A$ offers a unique 
possibility to search for heavy Higgs bosons not accessible elsewhere. The 
photons are generated by Compton back-scattering of laser light so that almost 
the entire energy of the electrons/positrons at a Linear Collider can be 
transferred to the photons \cite{ginzburg}, with luminosities of about one 
third of the $e^+e^-$ luminosity in the high-energy regime \cite{telnov}.

\begin{figure}[h]
\begin{center}
\epsfig{figure=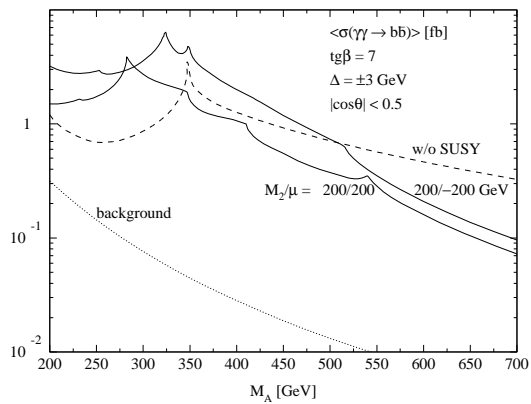,bbllx=50,bblly=221,bburx=566,bbury=616,height=5.3cm,clip=}
\caption{$H,A$ production cross sections in 
$\gamma\gamma$ collisions as a function of $M_A$ with final decays into 
$b\bar{b}$ and the corresponding background cross section. The MSSM parameters 
have been chosen as $\tan\beta=7,M_2=\pm\mu=200$~GeV; the limit of vanishing 
SUSY-particle contributions is shown for comparison \cite{Muhlleitner:2001kw}.}
\label{gamhaprod}
\end{center}
\end{figure}

In Ref.~\cite{Muhlleitner:2001kw} the search for $H,A$ in $\gamma\gamma$ 
collisions with subsequent decay into $b\bar{b}$ was analysed taking into 
account the NLO corrections to the signal \cite{nlosignal}, background 
\cite{nlobkg} and interference process. To enhance the signal to background 
ratio slim two-jet configurations have been selected in the final state, the 
incoming $e^\pm e^-$ beams have been chosen polarized and a cut on the 
scattering angle of the $b$-quark, $\theta$, has been applied. The maximum of 
the $\gamma\gamma$ luminosity has been tuned to $M_A$. The $b\bar{b}$ final 
states have been collected with a resolution in the invariant mass 
$M_A\pm \Delta$, $\Delta = 3$~GeV. (For more details see also \cite{thesis}.) 
In $\gamma\gamma$ fusion the mass reach can thus be extended to $\sim 80$\% of 
the total $e^+e^-$ energy, {\it i.e.} in the first phase of the ILC $H,A$ 
bosons can be discovered up to masses of about 400~GeV, and up to 800 GeV in 
the second phase, for medium values of $\tan\beta$, as can be inferred from 
Fig.~\ref{gamhaprod}. A detailed study taking into account all relevant 
theoretical and experimental issues has shown that the cross section can be 
determined with a statistical precision of 10\% and better \cite{piotr}.
\vspace*{-0.2cm}

The analogous analysis \cite{piotrsm} for the SM Higgs boson production in 
$\gamma\gamma$ fusion \cite{jikia} concludes that the partial width 
$\Gamma (H\to \gamma\gamma)$ can be extracted with 2\% accuracy for 
$M_H=120$~GeV. This provides a sensitivity to new charged particles running in 
the loop-induced $H\gamma\gamma$ coupling.

\subsection{Determination of $\tan\beta$ in $\tau^+\tau^-$ fusion}
\begin{figure}[h]
\begin{center}
\epsfig{figure=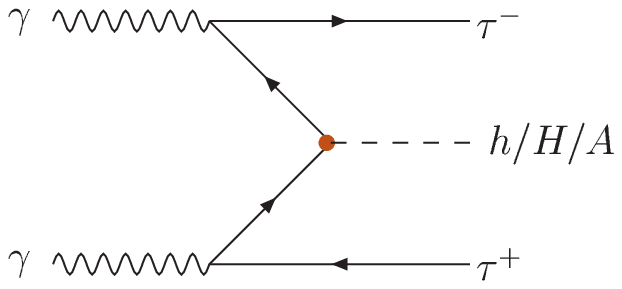,bbllx=238,bblly=614,bburx=420,bbury=705,width=3.7cm,clip=}
\hspace*{0.6cm}
\epsfig{figure=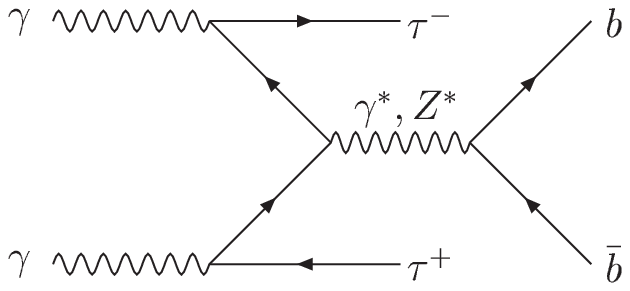,bbllx=238,bblly=614,bburx=420,bbury=705,width=3.6cm,clip=}
\hspace*{0.6cm}
\epsfig{figure=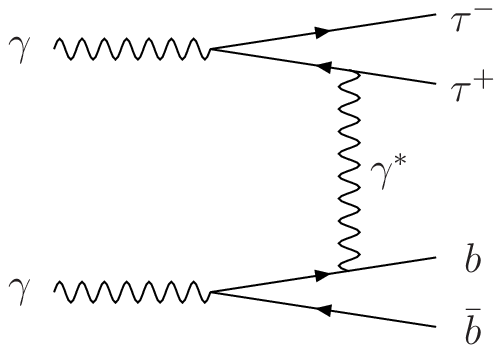,bbllx=238,bblly=614,bburx=420,bbury=714,width=3.6cm,clip=}
\caption{(a) The signal process $\tau\tau$ fusion into $h,H,A$. (b) The 
annihilation and (c) the diffractive background process.}
\label{taufus}
\end{center}
\vspace*{-0.2cm}
\end{figure}

Since the measurement of the important mixing parameter $\tan\beta$ is a 
difficult task and expected accuracies at the LHC and the ILC are at the order 
of 10\%, any additional method for its determination is valuable. The 
$\tau\tau$ fusion to $h,H,A$ at a PLC \cite{taufuspaper}, provides a promising 
channel and is based on the two-step process, {\it cf.} Fig.~\ref{taufus}a,
\beq
\gamma\gamma \to (\tau^+ \tau^-) + (\tau^+ \tau^-) \to \tau^+ \tau^- 
+ h/H/A
\eeq
For the large $\tan\beta$ case studied in \cite{taufuspaper}, 80 to 90\% of 
the Higgs bosons decay into a $b$ quark pair so that the final state consists 
of a pair of $\tau$'s and resonant $b$ quark jets. The couplings of 
$h,H,A$ to $\tau$ pairs being of the order of $\tan\beta$ 
\begin{figure}[h]
\begin{center}
\epsfig{figure=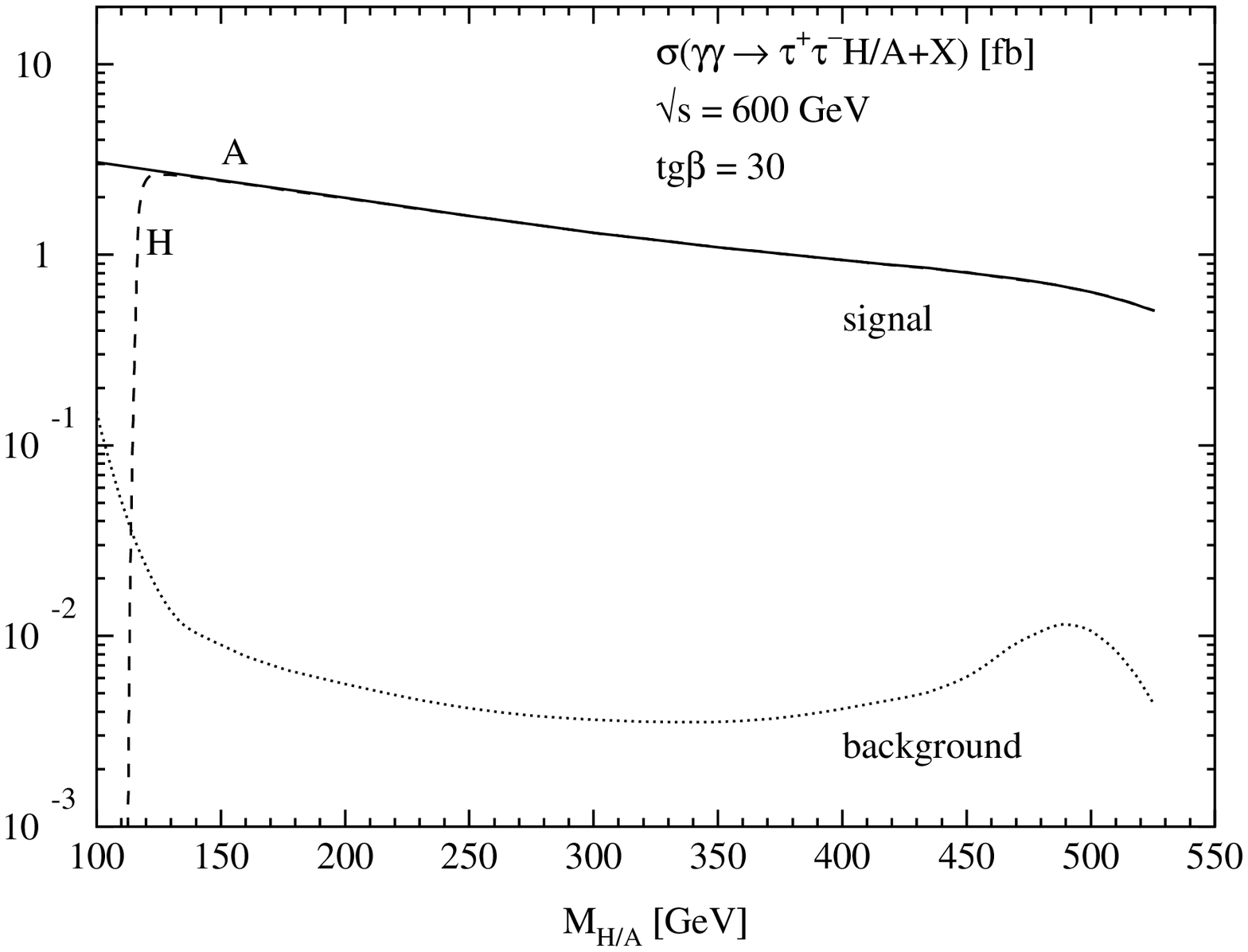,bbllx=54,bblly=223,bburx=564,bbury=617,height=4.6cm,clip=}
\hspace*{0.3cm}
\epsfig{figure=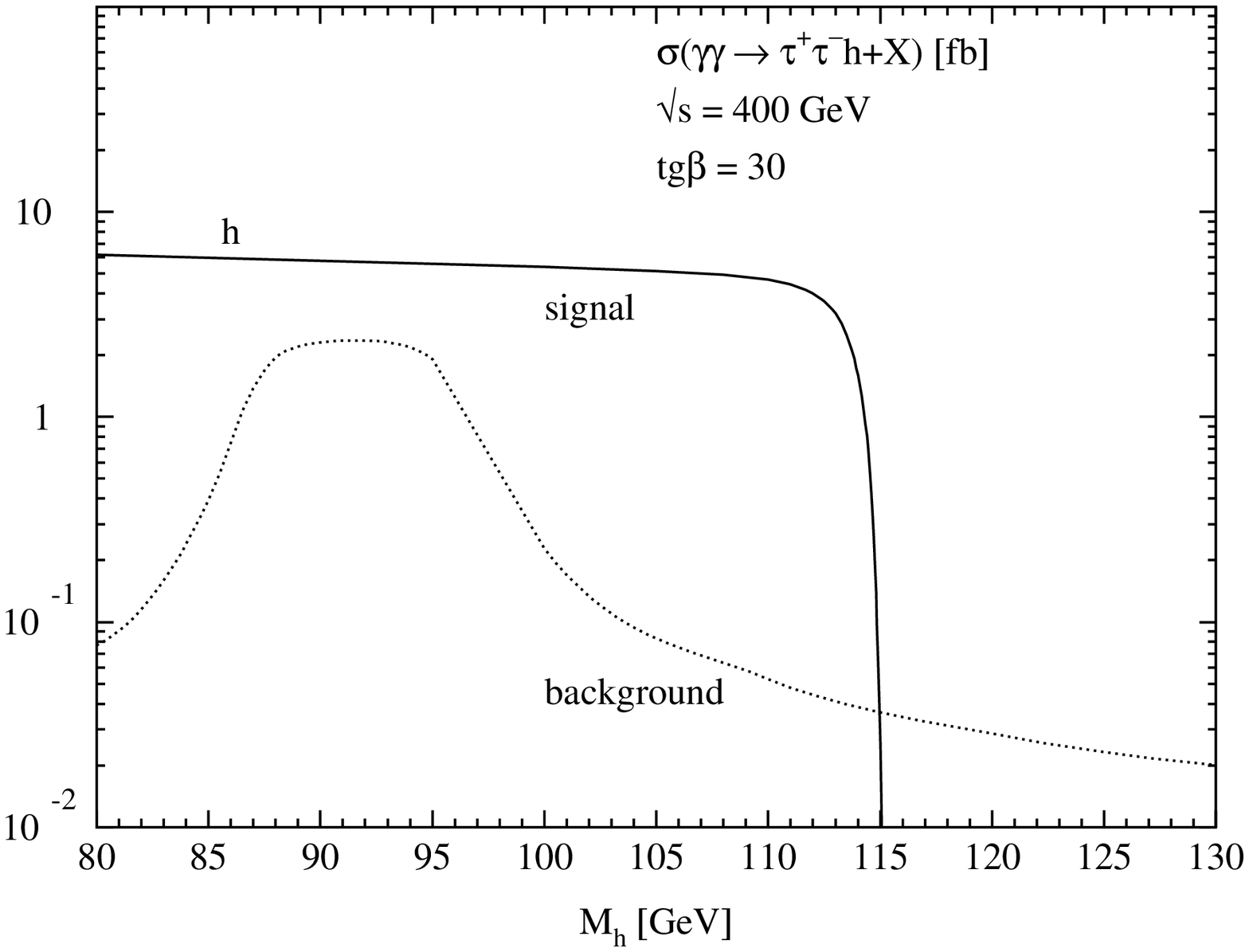,bbllx=50,bblly=223,bburx=566,bbury=617,height=4.6cm,clip=}
\caption{The $\tau\tau$ fusion into $H/A$ (left) and $h$ (right) for $\tan\beta=30$ compared to the background process. Cuts as specified in \cite{taufuspaper}. $\sqrt{s}$ denotes the $\gamma\gamma$ collider c.m. energy, {\it i.e.} 80\% of the $e^\pm e^-$ LC energy.}
\label{restaufus}
\end{center}
\end{figure}
\vspace*{-0.2cm}
(if $M_A$ is sufficiently light in the $h$ case) \cite{mondragon}, the signal 
process is enhanced for large $\tan\beta$ values. The main background channels 
are $\tau^+ \tau^-$ annihilation into a $b$-quark pair mediated by virtual 
$\gamma/Z$ exchanges (Fig.~\ref{taufus}b) and thus suppressed by the 
electroweak coupling, and diffractive $\gamma\gamma\to (\tau^+\tau^-)
(b\bar{b})$ events (Fig.~\ref{taufus}c), which is suppressed by choosing 
proper cuts. Results of the numerical analysis taking into account the full 
set of signal and background diagrams are shown in Fig.~\ref{restaufus}. 
Assuming standard design parameters of a PLC an error $\Delta\tan\beta\sim 1$, 
uniformly for $\tan\beta\gsim 10$, may be expected, improving on complementary 
measurements at the LHC and ILC.
%
\section{Summary}
At a future ILC the SM and MSSM Higgs bosons are accessible up to the 
kinematical limit independent of their decay properties. The precision on the 
masses is ${\cal O}(1$\%) and better. The PLC provides the unique possibility 
to discover heavy MSSM $H,A$ bosons in a wedge centered around medium 
$\tan\beta$ values, not accessible elsewhere, and complements the Higgs boson 
search at the ILC. Furthermore, the important mixing parameter $\tan\beta$ can 
be extracted in photon collisions with a statistical accuracy of 10\% and 
better for large values of this parameter. The PLC can thus be considered a 
valuable complement to the $e^\pm e^-$ mode of a future ILC.
%

\end{document}